\documentclass[fleqn,10pt]{wlscirep}
\usepackage[utf8]{inputenc}
\usepackage[T1]{fontenc}
\usepackage{subcaption}
\usepackage[ruled,vlined]{algorithm2e}

\title{Tangled Worldview Model of Opinion Dynamics}

\author[1,*]{Hardik Rajpal}
\author[1,2]{Fernando Rosas}
\author[1,3]{Henrik Jeldtoft Jensen}
\affil[1]{Centre for Complexity Science and Department of Mathematics, Imperial College London, South Kensington, London SW7 2AZ, UK}
\affil[2]{Department of Electrical and Electronic Engineering, Imperial College London, South Kensington, London SW7 2AZ, UK}
\affil[3]{Institute of Innovative Research, Tokyo Institute of Technology, 4259, Nagatsuta-cho, Yokohama 226-8502, Japan}

\affil[*]{Correspondance: h.rajpal15@imperial.ac.uk}


\begin{abstract}
We study the joint evolution of worldviews by proposing a model of opinion dynamics, which is inspired in notions from evolutionary ecology. Agents update their opinion on a specific issue based on their propensity to change -- asserted by the social neighbours -- weighted by their mutual similarity on other issues. Agents are, therefore, more influenced by neighbours with similar worldviews (set of opinions on various issues), resulting in a complex co-evolution of each opinion. Simulations show that the worldview evolution exhibits events of intermittent polarization when the social network is scale-free. This, in turn, trigger extreme crashes and surges in the popularity of various opinions. Using the proposed model, we highlight the role of network structure, bounded rationality of agents, and the role of key influential agents in causing polarization and intermittent reformation of worldviews on scale-free networks. 
\end{abstract}
\begin{document}

\flushbottom
\maketitle
\thispagestyle{empty}

\section{Introduction}
\label{intro}

\subsection{Motivation}

\textit{Weltanschauung}, usually translated from German as "worldview", is a concept popularised by the philosopher G.W.F. Hegel that refers to the fundamental cognitive orientation of an individual or society; i.e., the set of statements that are assumed consciously or unconsciously about the basic make-up of the world~\cite{sire}\cite{naugle_2002}. The worldview and experiences of different individuals affect each other and, interestingly, co-evolve together. In effect, worldviews drive human behaviour generating experiences that, in turn, are capable of triggering mutations in the worldview itself. Moreover, interactions between individuals with heterogeneous worldviews can foster significant mutations in their corresponding inclinations. Therefore, human interaction has always been a key driver of worldview evolution. 

The evolution of worldviews can exhibit fascinating complexity, as seen for e.g. in the evolution of political preferences. In effect, history shows that political worldviews usually exhibit a life cycle similar to biological organisms: they emerge, maybe become dominant, only for becoming extinct later \cite{spengler}. However, recent trends in liberal democracies have shown that political worldviews are sometimes capable of reviving after being inactive for a long time \cite{schwarz_2018}. Moreover, opposing worldviews can coexist and even reinforce each other, forming a fragile but persistent balance. This phenomenon, known as \textit{political polarization}, can lock worldview dynamics until the balance is abruptly disturbed, under events that may be seen very unexpected to external observers \cite{layman_2006}. Eminent cases of these ``surprising events" include the results of the US presidential election of 2016 and the Brexit referendum in the UK, which suddenly switched the balance in the worldwide political arena. 

It is tempting to dismiss abrupt changes in worldviews of societies as rare, idiosyncratic events, just as the spread of plague or mass extinctions in the $19^{th}$ century. On this point, it is interesting to note how the development of epidemiology changed our views on such subjects by enabling the prediction of sudden spreads of diseases\cite{scarpino2016}. Similarly, climate catastrophes like cyclones and floods were first considered to be aberrations, but nowadays advances in meteorology have uncovered non-linear dependencies that can be leveraged to understand these phenomena\cite{bathiany2018}. In a similar fashion, we propose that these sudden transitions in opinions can be regarded as indicators of a non-linear nature of worldviews' dynamics. This paper is an exploration of this rationale.

\subsection{State of the art}
 \label{sec:soa}

Most of the literature that studies polarization dynamics has been focused on simulation models (see e.g. \cite{Axelrod,starnini,hegselmann_krause}). These models and their variations have emphasized how homophily and confirmation bias can lead to global polarization.  
These studies find segregation of like-minded people in 2D space, on which agents interact with their spatial neighbours. However, the role of "space" and how it affect the herding dynamics in causing polarization is not well-understood. 

Considering that social media technologies have elevated the reach of individual influence from physical to the internet, modern approaches tend to place agents not in 2D space but rather connect them via various network topologies.The effect of various topologies was studied in \cite{amblard_deffuant_2004}, where they show that agents converge to one opinion above a critical value of connectivity, and this critical value decreases when the number of random links are increased over an initially regular network. Opinion dynamics have also been studied in models where the network structure changes at each time-step~\cite{baldassarri_bearman}. However, existing results suggest that it is difficult to isolate the role of topology in these scenarios due to the constant variations. 

From a modelling perspective, opinion dynamic models have evolved from simplistic contagion dynamics\cite{clifford_sudbury_1973,galam_2002} to sophisticated decision making agent based models\cite{stauffer_2002,deffuant_2000,hegselmann_krause}. Also, while most models focuses on the evolution of a single opinion, some body of work have explored the dynamics of collections of opinions. After the early works on cultural segregation reported in~\cite{Axelrod}, researchers have used the features of multi-opinion models to study the evolution of social power~\cite{jia}, clustering transitions~\cite{laguna} , and convergence with bounded confidence\cite{alaali}.

Most popular opinion dynamics models and consensus algorithms are such that their evolution tends to reach a single stationary steady state after some time. However, as observed in reality opinions usually don't reach a static equilibrium. Rather they follow a metastable behaviour and can exhibit punctuated equilibrium in time. This non stationary behaviour is also observed in the "noisy-voter model"\cite{Granovsky,Carro}, where the opinion change is governed by two driving parameters: Herding and Randomness. The model shows weak convergence for very small values of randomness. The added randomness keeps the system non-stationary. Although the model is able to reproduce some significant results, the added randomness acts as an extraneous parameter.

\subsection{Contribution and manuscript structure}

In this article we study worldview dynamics motivated by the ideas from evolutionary ecology. Worldviews are seen as an encoding of the basic presupposition of a belief systems, being analogous to  genes that encode biological signatures of specific species. Worldviews --- just like species --- emerge, dominate, and eventually go extinct. Moreover, just as new species are formed by mutations and interactions of existing species, worldviews emerge out of cross-cultural exchange. Finally, mass extinctions take place occasionally within ecosystems that eliminate existing species giving way to the emergence of new ones, which is analogous to sudden worldview shifts. 

The mentioned properties of evolutionary biology are captured by the well-known \textit{Tangled Nature Model}~\cite{jensen_2018}. This model, first presented in~\cite{Kim} as a simple agent based model of species-species interaction, is successful in showing how interdependent competition and mutations can lead to mass extinctions, which are followed by a chaotic phase of reformation that leads to a new state of meta-stable species. 

In this work we leverage this model to propose the \textit{Tangled Worldview Model of Opinion Dynamics}, in which agent's worldview is described by a binary vector and their dynamics is driven by a directed network topology. We use this model to develop our understanding of the role of topology in the formation of opinion's polarization, and also in the occurrence of abrupt global changes. Our model shows metastable evolution of worldviews punctuated by sudden abrupt transitions. Depending upon the network topology, these metastable states exhibit polarization or consensus.

The main findings reported from this model are the following:

\begin{enumerate}
\item Network topologies with just a few cycles do not foster polarization, while multiple cycles act as echo-chambers and reinforce polarization in agents with herding behaviour.
\item An important driver of sudden global reformations are changes in opinion of key agents in social network. These events trigger a cascade of opinion changes on various issues and a brief period of abrupt reformation, which allow new worldviews to emerge.
\item Bounded rational agents self-organize into consensus states on topologies with low cycles. However, on networks with many cycles even highly rational agents self-organize into polarized states.
\item By focusing on the popularity of a particular tagged opinion, the model exhibits noisy voter model-like dynamics (c.f Section~\ref{sec:soa}).  
\end{enumerate}

The rest of the article is structured as follows. First, Section~\ref{sec:2} introduces our proposed model. Then, Section~\ref{sec:3} presents our main findings from explorations on the model. Finally, Section~\ref{sec:4} summarises our main conclusions and points to future research.

\section{The model}
\label{sec:2}

This section introduces our model for worldview dynamics. In the sequel, first Section~\ref{sec:2.1} introduces some basic definitions, and  Section~\ref{sec:2.2} discusses the switching strategy used by the agents. Then the implementation of this strategy is explained with the help of an example in Section \ref{sec:2.3}. Finally, in the following section~\ref{sec:2.4}, structure of the considered social influence network is discussed along with the algorithm to generate it.

\subsection{Model description}
\label{sec:2.1}

Let us consider a system composed by $N$ social agents, where the internal state of the $i$-th agent corresponds to the binary vector $(O_1^i,\dots,O_K^i)$, with $O_j^i\in\{0,1\}$ for all $j=1\dots K$ . This vector represents the agent's worldview: the $K$ entries represents a particular subject/issue and its value (0 or 1) represents the position (for/against) of the agent on that topic. Note that $K$ is the number of considered issues within the worldview.

The social influence that agents have over each other is encoded into a social network. Social influence is generally asymmetric (contrasting with other social interactions like friendships, business etc) and heterogeneous. Hence we assume that the network of social influences to be represented by the adjacency matrix $A_{i,j}$ and is weighted by the weights $C_{i,j} \in [0,1]$ to introduce heterogeneity in the social influence network.

Agents interact with each other in this network, and change their opinions as a consequence of those interactions. This interaction includes, an agent switching its opinion on a particular issue based on various factors. In our model agents update their worldviews asynchronously. At each time step one agent is selected randomly, and it selects at random an opinion to update. In the sequel we consider the major factor that drive their dynamics.

\subsubsection{Affinity}
This quantity explores the effect of confirmation bias, i.e. the tendency of the agent to favour the opinion of those who agree with pre-existing opinions of the agents on other issues. The affinity between agent $i$ and $j$, denoted by $F_{i,j}$,is a quantification of similarity between the worldviews of any two agents say $i$ and $j$. There can be many ways to quantify this but for the purpose of this study we simply take the ratio of total number of issues with same opinion to the total number of issues. 

\begin{equation}
    F_{i,j} = \frac{1}{K} \sum_{n=0}^{K}\delta(O_n^i , O_n^j)
\end{equation}

The $\delta$ function in the previous equation is defined as follows, 

\begin{equation*}
\delta(O_n^i , O_n^j) =
    \begin{cases}
        1,& if\hspace{0.5cm} O_n^i = O_n^j \\
        0,& if\hspace{0.5cm} O_n^i \neq O_n^j
    \end{cases}
\end{equation*}

\subsubsection{Propensity and conformity} 

The propensity to change opinion on issue $k$ by agent $i$ , denoted by $P_{i,k}$. In our model this metric is calculated as
\begin{equation}
\label{Eq:2}
P_{i,k} =
    \begin{cases}
        \frac{\sum_{j=0}^{N} C_{j,i} F_{j,i} (1 - \delta(O_k^i , O_k^j))}{\sum_{j=0}^{N} A_{j,i} (1 - \delta(O_k^i , O_k^j))},& if\hspace{0.5cm} \sum_{j=0}^{N} A_{j,i} \neq 0 \\
        0,& if\hspace{0.5cm} \sum_{j=0}^{N} A_{j,i} = 0
    \end{cases} 
\end{equation}

This is a measure of average disagreement of agent $i$, with all those $j$ who influence it ($C_{j,i} \neq 0$), on issue $k$ weighted by the affinity($F_{i,j}$) of their worldviews. In the above equations, $1 - \delta(O_{i,k} , O_{j,k})$ gives the value $1$ every time when $O_{i,k} \neq O_{j,k}$. This in essence quantifies the weighted disagreement. $A$ refers to the adjacency matrix and $C$ refers to the weighted adjacency matrix of the network structure which is mentioned in the previous subsection. 

Whereas, conformity ($T_{i}$) is the measure of social conformity agent $i$ enjoys with those who are influenced by it. This metric is calculated as follows:

\begin{equation}
\label{Eq:3}
T_{i} =
    \begin{cases}
        \frac{\sum_{j=0}^{N} C_{i,j} F_{i,j}}{\sum_{j=0}^{N} A_{i,j}} ,& if\hspace{0.5cm} \sum_{j=0}^{N} A_{i,j} \neq 0 \\
        0 ,& if\hspace{0.5cm} \sum_{j=0}^{N} A_{i,j} = 0
    \end{cases}
\end{equation}

Here we quantify conformity as the average influence $i$ has on $j$ ($C_{i,j}$) weighted by the Affinity ($F_{i,j}$) between $i$ and $j$.  

\subsubsection{Social Fitness}

We introduce \textit{Social Fitness}, a variable denoted by $H_{i,k}$, to quantify the net social conformity enjoyed by an agent $i$ on a particular issue $k$. The social fitness is defined as the difference between the \textit{"conformity"} and \textit{"propensity"}. In essence, social fitness is a measure of difference between the approval of agent's followers and those who influence it. Agents therefore are inclined to change opinion on any given issue in order to maximize social fitness. Therefore, a lower value of social fitness warrants an opinion change on that particular issue.

\begin{equation}
\label{Eq:1}
H_{i,k} = T_{i} - P_{i,k}
\end{equation}

\subsection{Switching strategy}
\label{sec:2.2}
At each time step an agent is selected completely at random and the agent chooses one of the issue at random. The agent then analyses its social fitness on the chosen issue, to consider a change of opinion. The agents follow a probabilistic switching strategy. This strategy also provides a leeway to incorporate the rationality of agents. In principle, a completely rational agent will definitely switch on any issue that has negative value of social fitness. However, bounded rational agents might overestimate or underestimate this value. This error then leads to agents switching on issues they enjoy positive social fitness as well as keep opinions with negative social fitness with some probability. 
Therefore, the probability ($s_{i,k}$ of agent $i$ switching its opinion on issue $k$ depends upon the social fitness ($H_{i,k}$) and the rationality parameter ($\alpha$) as,

\begin{equation}
\label{Eq:4}
s_{i,k} = \frac{1}{1 + \exp(\alpha_{\text{eff}}H_{i,k})}
\qquad\qquad
\text{with}\quad\alpha_{eff} = T_{i} \alpha
\end{equation}

\begin{figure}
    \centering
    \includegraphics[width=0.5\textwidth]{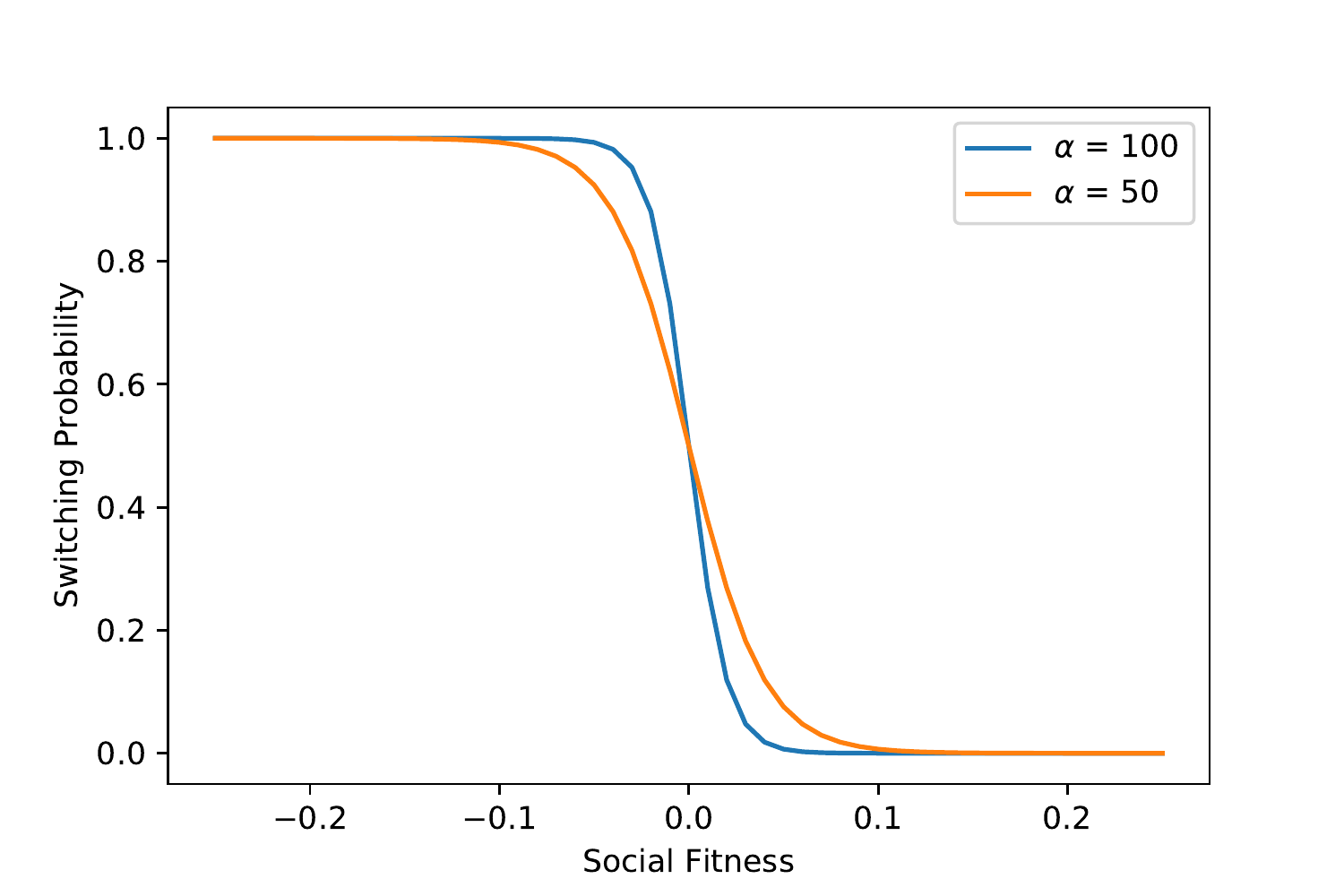}
    \caption{Effect of the rationality parameter ($\alpha$) on switching probability. It can be observed that for lower $\alpha$ probability of switching decreases for slight negative values of social fitness and subsequently increases for slight positive values of social fitness.}
    \label{fig:rational}
\end{figure}

Here the defined $\alpha_\text{eff}$ in Equation (\ref{Eq:4}), is an adaptive rationality parameter which is higher for agents with higher conformity. Whereas, $\alpha$ represents the rationality parameter of the agents or the ability of the agents to process the available information. For very high $\alpha$, the switching probability essentially becomes a step function about 0, i.e. the agents switch almost certainly even for small negative values of social fitness and almost never for positive values. However, as seen in Figure (\ref{fig:rational}) , for lower values of $\alpha$, the slope of the distribution around '0' becomes less steep. This allows the agent to switch with a finite probability for some positive values of social fitness as well as not switch at times for some small negative values. This inability to optimally process the available information allows for a bounded rational behaviour\cite{Ortega} which leads to more variability in behaviour of the agents. The adaptive $\alpha_\text{eff}$ accounts for the fact that agents with higher social conformity react more efficiently to the social pressure. Mirta et. al. \cite{Mirta} refer to this parameter as a measure of efficiency of flow of information among the agents.

\subsection{Example}
\label{sec:2.3}
Let us illustrate how the model works with an example. 
\begin{figure}
    \centering
    \includegraphics[width=0.4\textwidth]{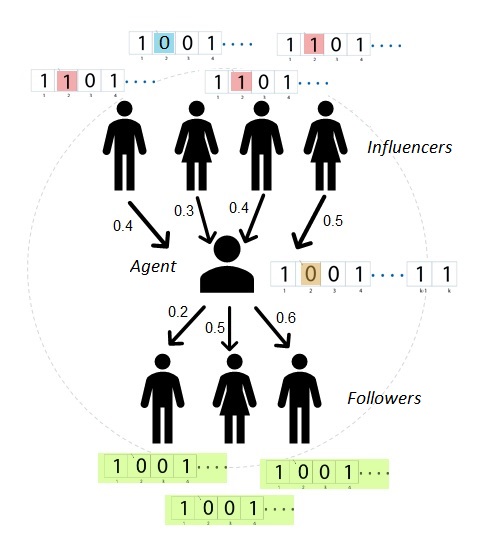}
    \caption{Schematic representation of the model. The agent considers the propensity exerted by the influencers and the conformity enjoyed with the followers, when deciding to switch on issue number 2. Three of the influencers disagree with the agent on issue number 2 (represented by red), while only one agrees (represented by blue). All the followers of the agent (those influenced by the agent) agree with the agent on all given issues (represented in green).}
    \label{fig:Illus}
\end{figure}

At any particular time step an agent is selected, and the network structure around the agent looks as presented in Figure~\ref{fig:Illus}. The agent chooses (randomly) to consider his opinion about issue number 2 (highlighted in the Figure in yellow). As illustrated in the figure, 3 agents disagree with the agent on this issue (highlighted in red). For simplicity, we assume that this is the only issue these agents disagree upon. 

The three disagreeing agents exert a social pressure on the agent to change its opinion on this issue. This social pressure corresponds to the propensity, and is calculated according to \eqref{Eq:2} as
\begin{equation}
    P_{i,2} = \frac{0.4\times\frac{K-1}{K} + 0.4\times\frac{K-1}{K} + 0.5\times\frac{K-1}{K}}{3}.
\end{equation}
For example, if $K = 8$ then $P_{i,2} \approx 0.4$

The agent also enjoys social conformity with those who are influenced by this agent. For simplicity, let us assume that all the three agents agree with the agent under consideration on all issues. Thereby, these agents contribute to the social pressure on the agent to keep his opinion on the issue unchanged. This kind of pressure corresponds to conformity and can be calculated from \eqref{Eq:3} as
\begin{equation}
    T_{i,2} = \frac{0.2\times\frac{K}{K} + 0.5\times\frac{K}{K} + 0.6\times\frac{K}{K}}{3}.
\end{equation}
For $K = 8$ then $T_{i,2} \approx 0.43$ 

Finally, the social fitness of the agent on this issue is $H_{i,2} = 0.025$. Therefore, the switching probability can be calculated as from \eqref{Eq:4} to be
\begin{equation}
    s_{i,2} = \frac{1}{1 + \exp(\alpha\times T_{i,2}H_{i,2})}.
\end{equation}

In particular, if $\alpha = 100$ then the switching probability is $s_{i,2} \approx 0.26$. Therefore the agent will switch its opinion with a probability of $0.26$ and the same procedure is repeated for other agents.

\subsection{Network Structure}
\label{sec:2.4}
The diverse set of human interactions drive various emergent consequences, therefore it is imperative to identify the interaction most relevant to the problem at hand in order to model it mathematically. Literature from social psychology\cite{Kelman, Mehdi, Chacoma} assert the fact that social influence plays a leading role in causing opinion change. Therefore for the purpose of this study we consider social influence as the primary cause for an agent changing its opinion. In the paragraphs below we describe the simulated social-influence networks we will be working with. Although there is no clear empirical evidence of what the topology of human social influence networks is like, it is clear that human societies are heterogeneous and hierarchical. Some agents are more competent in certain abilities and hence yield more influence over others. This suggests that the distribution of social influence in social structures follow a Pareto distribution. This assumption is backed by many studies of online social networks \cite{Mislove, Anghel}. Motivated by this discussion, the network of social influence can be empirically expected to be a directed scale-free network. The following Algorithm \ref{algo} describes how such a network is generated for our analysis.\newline

\begin{algorithm}[H]
\label{algo}
\DontPrintSemicolon
\SetAlgoLined
\BlankLine
Initialize the matrices $C$ and $A$ to be null matrices of shape $N \times N$.
\For{$i=1,\dots,N$}{
    \eIf{$i \leq n_0$}{ Set $A_{i,j} = 1 \hspace{1cm} \forall j \in {1,\dots,n_0}$\;}{
    \For{$k=1,\dots,n_0$}{
        Select $j$ randomly such that, $j \in {1,\dots,i-1}$\ and $A_{j,i} \neq 1$\;
        $i$ draws a link, $j \rightarrow i$ with a probability $\frac{\sum_{m=0}^{N} A_{j,m}}{\sum_{n=0}^{N} \sum_{m=0}^{N} A_{n,m}}$\;
       }
     Set the self link, $A_{i,i} = 1$.\;
     To create a back link (cycle), with probability $c$, draw $r$ uniformly at random between $[0,1]$.\;
     \If{$r<c$}{
     Select $j \in {1,\dots,i-1}$ and $A_{i,j} \neq 1$.\;
     Set $A_{i,j} = 1$.\;
     }
    }
    \For{$j=1,\dots,N$}{
    Draw a random number $w$ from a symmetric beta distribution ($\alpha = 5, \beta = 5$) between $[0,1]$\;
    Set, $C_{i,j} = w \times A_{i,j}$\;}
}
\caption{Algorithm for generating the Adjacency matrix for the social influence network}

\end{algorithm}
\vspace{1cm}
Note that initially the network has cycles and self loops only among the first 5 agents. More cycles are introduced before the final step in the proposed algorithm as node $i$ creates a influence link to random node $j (j<i)$ with some probability $c$. Now as the node $i$ introduces a "back" link, it automatically completes an influence cycle. This means that influence on any opinion expressed by agent $i$ travels back to the agent in a loop and reinforces that influence. In a social context these influence loops are analogous to echo chambers. The probability $c$ behaves as a parameter to control the number of such influence cycles.\newline
In order to introduce a certain degree of heterogeneity of social influences as each influence is different from the other, the adjacency matrix ($A$) is then weighted by weights drawn from a symmetric beta (5,5) distribution between [0,1]. The choice of this distribution is based on the assumption that extreme influences are less likely. The resulting weighted matrix is denoted by $C$.

\section{Results and analysis}
\label{sec:3}

\subsection{Method description}

For a detailed analysis in this model, a social network ($N = 100$) of highly rational ($\alpha = 100$) social agents is considered. The worldview of these agents comprises of opinions on $k = 8$ issues. The agents are updated asynchronously, i.e. at each time step an agent is selected at random and the agent selects an issue at random. Thereafter, the agent updates its opinion on the selected issue as described in the example above. We record the state of the system after every $N \times k$ time steps. The state of the system is identified by worldview IDs, i.e. the decimal conversion of the binary string of 'k' bits, for each agent. This enables us to study the time evolution of both worldviews and individual opinions. The worldview IDs are randomly initialized for the agents.  \newline

Simulations start by generating a social network as described in Algorithm (\ref{algo}). First, we obtain an adjacency matrix, $\textbf{A}$, ($100 \times 100$) and the weighted adjacency matrix $C$. We simulate the dynamics as discussed in the example in the previous section, on topologies with different back link probabilities ($c$).

\subsection{Polarization and cycles}

\begin{figure}
    \centering
    \includegraphics[width=0.7\textwidth]{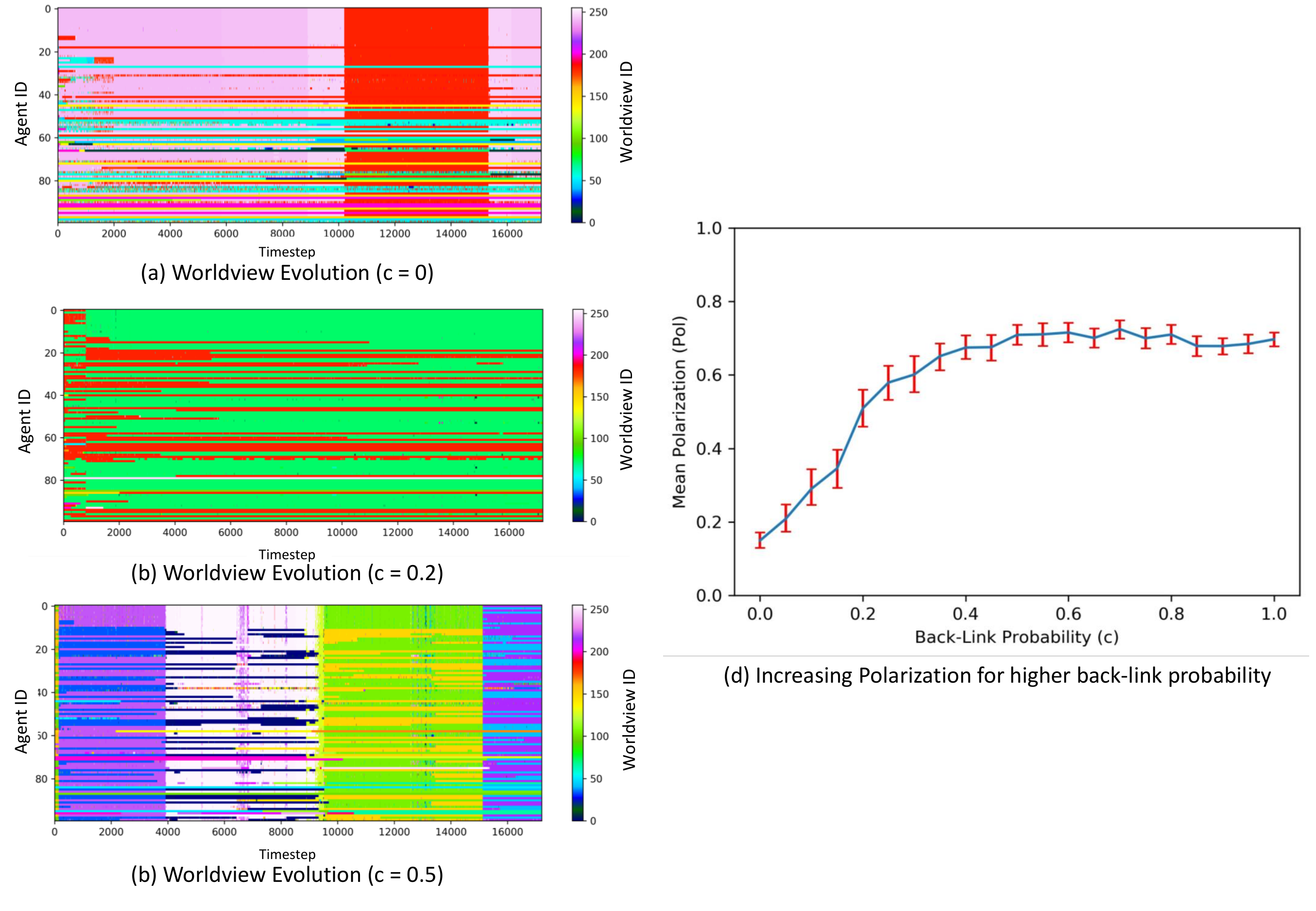}
    \caption{(a,b,c) Worldview evolution for 100 agents is observed for three different values of back-link probabilities (c = 0, 0.2, 0.5). The horizontal axis of the figure represents the time steps and the vertical axis represents individual agent IDs. The colours at each time-step indicate the Worldview ID of each agent. It is observed that polarization of worldviews increases with more back-links as agents self-organize into opposing worldviews. Some sudden transitions are also observed.\newline (d)The variation of polarization for various values of back-link probability (c). The error bars show the 95\% Bootstrap Confidence Intervals for the mean values obtained by 100 simulations for each value of 'c'.}
    \label{fig:polar}
\end{figure}

We first explore the effect of the number of cycles on the opinion dynamics. Figure \ref{fig:polar} (a,b,c) illustrates the typical behaviour for three conditions: $c = 0$ (No extra back-links), $c = 0.2$ (A fifth of the agents create a back-link) and $c = 0.5$ (Half of the agents create a back-link). As discussed before, back link probability ($c$) controls for the number of cycles in the social influence network.

In the first case, it can be observed that most of the agents ($60\%$) have the same worldview, and the opposing views are not necessarily the polar opposite of the dominant worldview. For instance, the most popular worldview at $T = 8000$ in this case is Worldview ID: 243 (Binary representation: '11110011') which is adopted by 60 agents. The next most popular worldview is worldview ID: 56 (Binary representation: '00111000') which is popular among only 11 agents. Also, these two worldviews disagree on some but not all the issues. 

However, for $c = 0.2$, some polarization emerges as a minority of agents hold a worldview which is completely opposite to the majority worldview. For example, in this case at $t = 8000$, the most popular worldview ID is 72 (Binary Representation: '01001000') which is popular among 70 out of the 100 agents, however there is a significant minority of 27 agents which follow the worldview ID: 183 (Binary Representation: '10110111'). Here the majority and the minority disagree on all the issues, representing a case of group polarization.

Finally in the case of $c = 0.5$, the majority weakens more and almost half of the population follows a worldview, while a significant opposition follows a completely opposing worldview. For example, in this cast at $t = 8000$, the majority of 57 agents believe in worldview ID 254 (Binary Representation: '11111110') and the significant minority holds the polar opposite worldview 1 (Binary Representation: '00000001'). This observation suggests that the inclusion of cycles/back-links can indeed lead to this emergent global polarization. 

Therefore to measure this it is imperative to discuss a measure of polarization we use to quantify polarization at any given state of the system. Consider the state of the system at any time $t$ is given by $n_1 : w_1, n_2 : w_2 ....$, i.e. $n_i$ agents with worldviews $w_i$. Suppose the most popular worldview is $w_j$, followed by $n_j$ agents and the second most popular worldview is $w_k$, followed by $n_k$ agents. We define
\begin{equation}
    Pol = \frac{d(w_k,w_j)\times n_k }{n_j}
    \newline
    Where,\newline
     d(w_k,w_j) = \frac{1}{K}  \sum_{n=0}^{K}(1 - \delta(O_n^k , O_n^j))
\end{equation}

as a metric of polarization. Above, the function $d(w_k,w_j)$ gives the hamming distance between the worldviews $k$ and $j$, and $N$ represents the total number of agents. For example, if the system gets into a state where half of the population has an opposite worldview as the other half, then $Pol = 1$. On the other hand, if system gets into a state where all the agents have the same worldview, $Pol = 0$. 

The average polarization for 17000 timesteps over 100 simulations for various values of back-link probabilities $c$ is calculated. Figure \ref{fig:polar} (d) shows a slow but continuous increase of polarization with the back link probability, confirming that cycles are associated with polarization. 

A back-link completes an influence cycle and hence creates an echo chamber. When the agents, their influencers and their followers, have the same worldview, the probability of anyone in the echo-chamber to change opinion an any issue becomes very small. 

In summary, polarization can be said to be the result of agents' trying to minimize propensity (or maximize social fitness). By recalling equation (\ref{Eq:2}), agents can do this in two ways: either by coming to the same worldview as all its neighbours (thus there is no disagreement on any issue), or by taking the completely opposite worldview, in which case the fractional similarity of the worldviews goes to zero and consequentially, the propensity. 

In the $c = 0$ case, agents initially get into a polarized state to minimize propensity. But the absence of any cycles (i.e. echo-chambers), eventually a dominant majority is always established. However, as soon as echo chambers are introduced, then agents tend to stay in polarized states. Transitions are triggered by rare changes of opinions by key agents, causing the system to reorganize in a different polarized state.

\subsection{Sudden transitions}
Simulations show sudden transitions of worldviews, which can be seen in Figure~\ref{fig:polar}(c) (3 transitions), and also in \ref{fig:polar}(a) (2 transitions). Note that these transitions are rare, as they occur when certain key agents (usually the ones with high influence) change their opinion on certain issue and thereafter triggering a cascade of opinion changes. After the cascade the system self-reorganizes in a different meta-stable state, which remains with minor fluctuations until another transition is triggered. This is related to the sigmoidal nature of the switching probability, as agents with very high social conformity and positive social fitness have a small yet non-zero probability of switching their opinion on any issue.  

\subsection{The effect of rationality}
\begin{figure}
  \centering
  \includegraphics[width=0.8\textwidth]{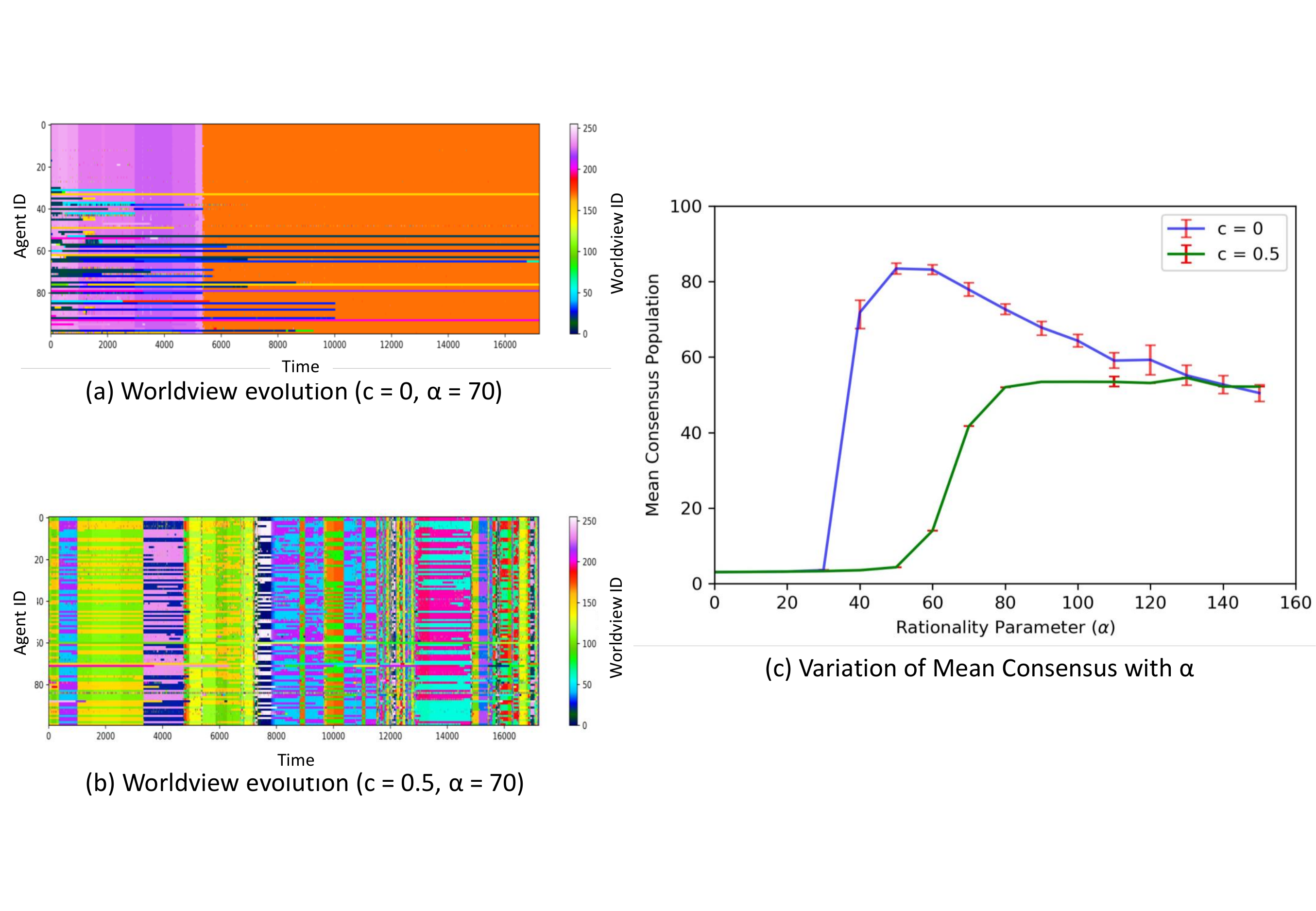}
  \caption{(a,b) Worldview evolution for agents with same rationality parameter on different topologies ($c = 0$ , $c = 0.5$). It is observed that More polarization and more transitions are observed in ($c = 0.5$) for lower value of rationality.\newline  (c) The effect of decreasing the rationality parameter on mean consensus population for 17000 time sweeps (100 simulations). Error bars represent the 95\% Bootstrap Confidence Intervals.} \label{temp}
\end{figure}

To study the role of rationality parameter ($\alpha$), we consider networks for the cases $c = 0$ and $c = 0.5$. The worldview evolution for both network topologies for $\alpha = 70$ is shown in the Figure (\ref{temp}). When comparing this with the worldview evolution corresponding to $\alpha = 100$ in Figure (\ref{fig:polar}), it can be observed that the consensus is enhanced for lower $\alpha$ for $c = 0$ but has an opposite effect for $c = 0.5$. To quantify this, we calculate the  average number of agents following the most popular worldview (Mean Consensus) for various values of efficiency parameter ($\alpha$), for over a 100 simulations in both cases. The Figure \ref{temp}(c), shows the variation of mean consensus population with the rationality parameter for both topologies. It is observed that for $c = 0$, the mean consensus overshoots half the size of the population, quickly for a lower value of critical rationality than for $c = 0.5$. However, for $c = 0.5$, the transition is more similar to that in Ising model\footnote{A mathematical model of ferromagnetism in Statistical Physics}, where the consensus quickly rises to the maximum value of half the population after the critical value of rationality.

\subsection{Dynamics of individual opinions}
\begin{figure}
    \centering
    \includegraphics[width =0.6\textwidth]{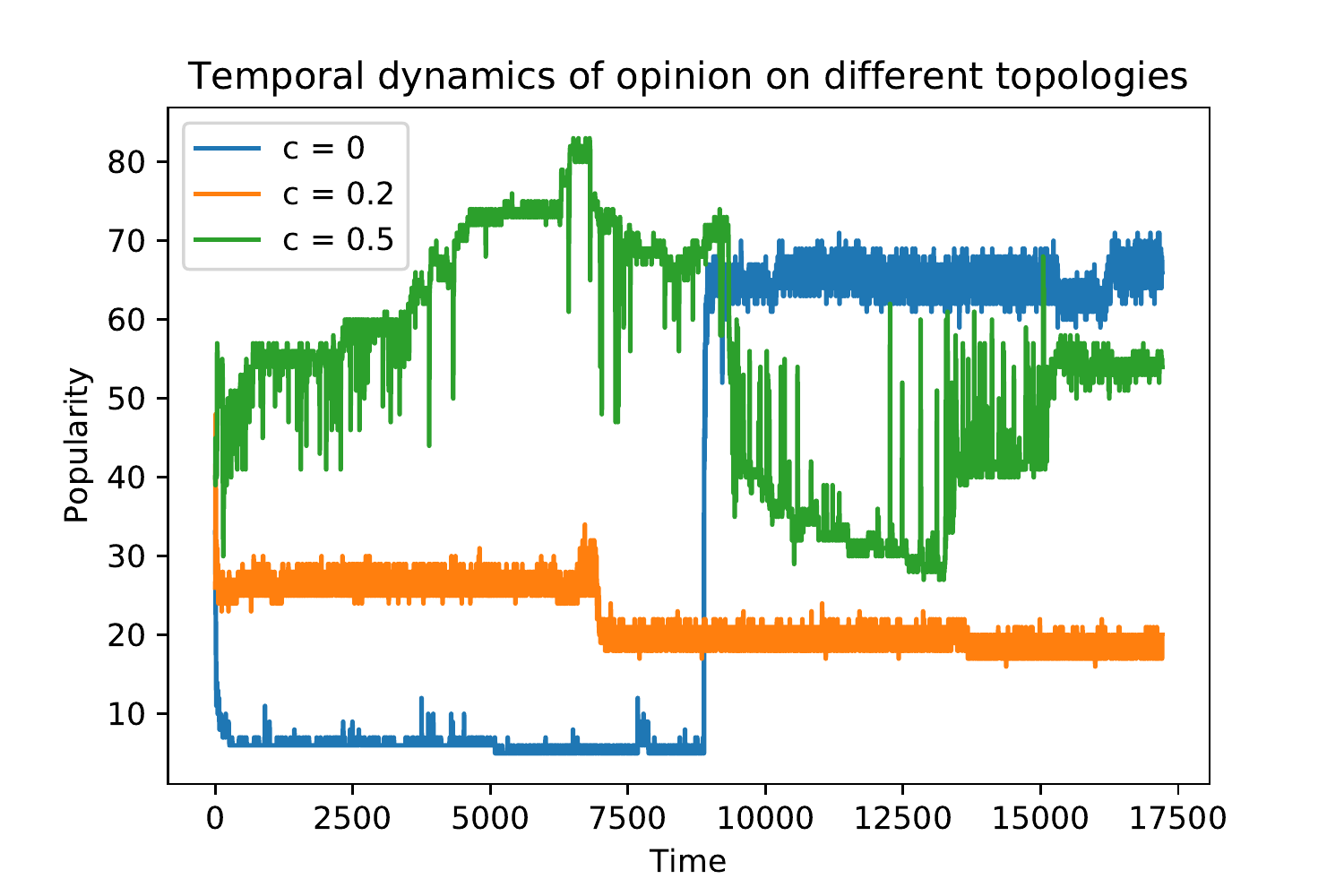}
    \caption{Time evolution of number of agents in favour of issue number 6. Sudden surges are seen during worldview transitions in consensus topologies (c = 0, c = 0.2). However, the variation of opinions is much more continuous, with flash crashes/surges, in polarized topology (c = 0.5)}\label{fig:OD}
\end{figure}

We now discuss the dynamics of opinions on individual issues. Figure~\ref{fig:OD} shows the dynamics of a particular opinion on issue '5' in the three discussed values of the cycle parameter (c). 

The dynamics in the case of consensus topologies ($c = 0$ and $c = 0.2$) shows that variation in popularity of opinions changes very slightly and sudden surges/crashes happen when there is a sudden transition as discussed in previous sections. However, in the case of polarized topology the popularity of opinion shows much more variability as well along with more flash crashes and surges. This volatility can be attributed to the frustration introduced by polarization, as some agents, which are influenced by both poles, tend to switch back and forth. 

On the other hand, in a consensus topology, where the other pole doesn't exist, the popularity doesn't change much during a meta-stable state. In the consensus case, one can observe aperiodic transitions between popular and unpopular phases if a long term evolution is considered, as illustrated in Figure \ref{fig:OD}. 

For certain parameters, this behaviour is concurrent with the behaviour observed in the noisy voter model\cite{Carro}, which successfully explained the volatility clustering in opinion models. Therefore, the proposed \textit{Tangled Worldview Model}, provides a much more general framework to study opinion dynamics.

\section{Conclusion}
\label{sec:4}
 We proposed a stochastic model for the dynamics and co-evolution of the worldview of agents that participate in a social network. This formalism successfully captures highly non-trivial macroscopic social phenomenon, including polarization, sudden worldview transitions, and the effect of rationality of agents. The observed dynamics of opinions shows a intermittent behaviour with sudden changes related to worldview transitions.

We discussed the role of network topology in generating and sustaining polarization, given the herding dynamics of the model. It was observed that social influence networks with cycles reinforce and facilitate polarization. Conversely, networks with less cycles and more tree like (hierarchical) structure facilitate consensus. This finding also provides a possible explanation for the tree like organizational structure observed in most organizations. Such a network structure facilitates the flow of command and propagation of values of the particular organization. This finding also helps in explaining the cultural segregation observed in \cite{Axelrod} and polarization observed in \cite{deffuant_2000,hegselmann_krause}. These models considered the social topology to be either an undirected 2D grid or completely connected. Both of these topologies facilitate in polarization/segregation as they have multiple cycles. Although there is no denying the fact that local preferential herding driven by homophily is one of the prime drivers of polarization. The findings of this paper suggest that cycles in social networks restrict the system in reaching complete consensus. However, if these cycles are reduced consensus becomes possible even in presence of preferential herding behaviour like homophily.

It is also observed that sudden transitions in prevailing worldviews are possible even when absolutely no changes in behaviour of the agents, or the network topology are considered. This finding exposes the crucial role of the network topology, and the relevance of certain key influential agents in the social structure. 
In particular, sudden transitions were found to be usually attributed to influential agents that change opinion on a particular issue and consequently trigger a cascade of opinions on all issues. This leads to an abrupt major rearrangement, which relaxes temporally into another meta-stable state. This sort of dynamics could explains the punctuated equilibrium of worldviews, as observed e.g. in real world politics (For example: Synchronized weakening of liberal democracies and resurgence of nationalism are such transitions observed within the last decade). Therefore, in the age of online social networks and echo chambers, it is imperative to take up the challenges associated with recommendation algorithms and choice architecture\footnote{the practice of influencing choice by changing the manner in which options are presented to people.\cite{thaler_sunstein_2009}} that can foster more harmonious online and offline social behaviour.

That said, there are a couple of limitations of the proposed model that need to be mentioned, which we plan to do explored in future work. Firstly, as only one of the human biases is taken into account -- namely, confirmation bias, polarization is the only emergent social phenomenon observed among the agents. Human behavioural choices are in general much more complex, which leads to wider variety of emergent processes. However, due to the effects of non-linearities, a separate study of each of these biases and the macroscopic behaviour it generates will underscore the roles of such biases in the overall dynamics. Secondly, polarization is capable of inducing  changes in the social network topology. Therefore, in many scenarios the network of social influence might change and evolve in time. It is therefore important to develop an understanding of the type rules of temporal evolution of social influence networks that can drive system into and out of polarized states.

\bibliographystyle{model1-num-names}
\bibliography{sample.bib}

\section*{Acknowledgements}
We thank Pedro A. M. Mediano for the useful discussions. We also thank Imperial College Research Computing Service,  DOI: 10.14469/hpc/2232 for the computing facilities. HR is supported by the Imperial College President's PhD Scholarship. FR was supported by the European Union's H2020 research and innovation programme, under the Marie Sk\l{}odowska-Curie grant agreement No. 702981.

\section*{Author contributions statement}
All authors contributed equally to the conceptualization of the model. HR conducted the simulations and analyzed the results. All authors contributed to the writing, and approved the final version of the manuscript.

\section*{Additional information}
\textbf{Competing interests}
The author(s) declare no competing interests.

\end{document}